\documentclass[twocolumn,showpacs,preprintnumbers,amsmath,amssymb,floatfix]{revtex4}
\usepackage{bm}

\newcommand{\nabb}{{\bm{\nabla}}}
\newcommand{\gtapprox}{{\raise.3ex\hbox{$>$\kern-.75em\lower1ex\hbox{$\sim$}}}}

\usepackage[dvips]{epsfig}

\usepackage{float}

\usepackage{amsmath}
\usepackage{amsfonts}

\newcommand{\begineq}[1]{\begin{equation}\label{#1}}
\newcommand{\eqend}{\end{equation}}

\begin{document}
\title{Smectic phases with cubic symmetry: the splay analog of the blue phase}

\author{B.A. DiDonna} 
\author{Randall D. Kamien}
\affiliation{Department of Physics and Astronomy, University of Pennsylvania, Philadelphia, PA 19104-6396, USA}

\begin{abstract}{
We report on a construction for smectic blue phases, which have quasi-long range smectic translational order as well as long range cubic or hexagonal order. Our proposed structures fill space with a combination of minimal surface patches and cylindrical tubes. We find that for the right range of material parameters, the favorable saddle-splay energy of these structures can stabilize them against  uniform layered structures.
}\end{abstract}
\maketitle

Liquid crystalline blue phases exhibit true three-dimensional, periodic orientational order.  Two of these phases possess cubic symmetry ($BP1$ and $BP2$) while the third ($BP3$) is thought to be an isotropic melt of double-twist cylinders \cite{spaghetti1,spaghetti2}.  Recently, new phases of matter have been
identified that possess the quasi-long range translational order of smectics~\cite{mhli} and,
at the same time, three-dimensional orientational order.  These three distinct smectic blue phases have been observed near the isotropic transition of these compounds: $BP_{sm}1$ has cubic symmetry, while $BP_{sm}2$ and $BP_{sm}3$ have hexagonal symmetry. The precise physical properties of these materials have been the study of intense investigation in recent years~\cite{pansubp1, pansubp2, pansubp3}. However, there is no obvious way to incorporate smectic ordering into the traditional double-twist tube blue phase ordering put forward by Sethna {\sl et al.}~\cite{sethna} for nematic blue phases. In general, since smectic ordering is incompatible with cubic symmetry, it is expected that any blue phase structure must include smectic defects as well as orientational defects. Though a model for double-twist cylinders with smectic order
has been proposed~\cite{kamien}, the simplest variant of that model is incompatible
with experimental details~\cite{pansubp2}.
In this letter, we propose a new scheme for constructing a smectic blue phase that fills space with continuous concentric layers with cubic symmetry.  For this construction there are elastic
energy costs arising from non-uniform layer spacing and layer bending as well as condensation energy costs arising from melted regions.  However,
we show that when the saddle-splay constant, $K_{24}$, is negative enough, these energy costs
can be compensated by the gain in Gaussian curvature energy in the surfaces.  

The key ingredient in our construction is the observation that saddle-splay and the Gaussian
curvature are identical~\cite{rmp} for layered systems with uniform spacing.  
The saddle splay energy of a director field ${\bf N}$ is~\cite{sethna}
\begin{equation}
F_{SS} = K_{24} \int d^3 x \,\nabb \cdot \left[ \left( {\bf N} \cdot \nabb \right) {\bf N} - {\bf N} \left( \nabb\cdot {\bf N} \right) \right],
\label{eq:ssd}
\end{equation}
where $K_{24}$ is a Frank constant. This term is normally neglected since it is a total derivative and thus, for most geometries, becomes an unimportant boundary term. However, defects introduce boundaries, and it is precisely the saddle-splay energy that stabilizes the standard nematic blue phase. Since that phase is riddled with defects, the saddle-splay boundary energy grows linearly with volume.  However, if space can be filled with surfaces with normal $\bf N$ then $F_{SS}$ has a different interpretation -- if we replace the $z$ integration
in (\ref{eq:ssd}) with an integration over a Lagrangian coordinate $n$, which labels
the surfaces, then 
\begin{eqnarray}\label{eq:ssad}
F_{SS} &=& -2K_{24} \int \!dn\!\int dxdy\,  \frac{dz}{dn}K_n(x,y)\\
&=& -2K_{24} \int \!dn\!\int dxdy\sqrt{g_n(x,y)}\,a_n(x,y)K_n(x,y)\nonumber
\end{eqnarray}
where $a_n(x,y)$ is spacing between layer $n$ and $n+1$ at $(x,y)$, 
$K_n$ is the Gaussian curvature of the $n^{\rm th}$ surface, $g_n$ is
the determinant of the two-dimensional surface metric, and $\sqrt{g_n}a_n=dz/dn$ follows
from conservation of volume.  Since the Gauss-Bonnet theorem~\cite{rmp} implies that for a surface
of genus $g$
the integrated Gaussian curvature is $4\pi(1-g)$, we expect that higher-genus surfaces are favored by the saddle-splay term.  Note that since $a_n(x,y)$ is not necessarily constant, the integral (\ref{eq:ssad}) is not simply topological.  Nonetheless this identification will aid us in our
choice of smectic structures.   We note that in our model, the saddle-splay is a measure
of the layer normals, not the nematic director.  When the nematic director follows the
layer normal these are, of course, equivalent.  However, in type-II smectics it is possible
for the saddle-splay of the director field to differ in its precise numerical value from the saddle-splay in the layers.

We must choose our structures in the context of the standard bulk free energy of a smectic liquid crystal:
\begin{equation}
\label{eq:smfe}
F_{Sm} = \int d^3x \, \left\{\frac{B}{4}\left[\left(\nabb\Phi\right)^2-1\right]^2 + {2K_1} H^2\right\}
\end{equation}
where the smectic density is $\rho\propto\cos\left(2\pi\Phi/a\right)$, $\Phi(x,y,z)$ is a phase field, $a$ is
the layer spacing, $B$ is the compression modulus, $K_1$ is the bend modulus, and $H=\frac{1}{2}\nabb\cdot{\bf N}$ is the
mean curvature of the layers.   Though we could capture the same physics by replacing $H$ with
$\frac{1}{2}\nabla^2\Phi$ we choose the former for computational convenience.

We note that there is an intrinsic frustration between the two terms in (\ref{eq:smfe}).  Consider a
family of 
layers ${\bf x}_n(\sigma,\tau)$ which are solutions to $\Phi\left({\bf x}_n(\sigma,\tau)\right) = n a$.  
If ${\bf N}=\nabb\Phi/\vert\nabb\Phi\vert$ is the normal to ${\bf x}_0$, then ${\bf x}_n = {\bf x}_0 + na{\bf N}$ is a solution with vanishing compression energy: ${\bf N}\cdot\nabb\Phi\left({\bf x}_n\right)=\frac{d}{d(an)}\Phi\left({\bf x}_0 + na{\bf N}\right) = 1$ and so $\vert\nabb\Phi\vert =1$.   
However, if $\kappa_{i,0}$ are the principal curvatures
of ${\bf x}_0$, then since these surfaces are uniformly spaced, $\kappa_{i,n} = \kappa_{i,0}/(1+na\kappa_{i,0})$ are the principal curvatures of
${\bf x}_n$.  Thus, if the mean curvature $H=\frac{1}{2}\left(\kappa_1 + \kappa_2\right)$ vanishes for some value of $n$, it will
not vanish for any other value of $n$ unless $\kappa_{i,n}=0$.  Conversely, if $H$ vanishes everywhere and $\kappa_{i,n}$ does not, then it
is impossible to have equally spaced layers.  Therefore it is impossible to have both the
bending and the compression terms vanish unless all the layers are flat.
Thus any ground state must necessarily be a compromise between non-uniform spacing and
bending.  

\begin{figure}

\center

\epsfig{file=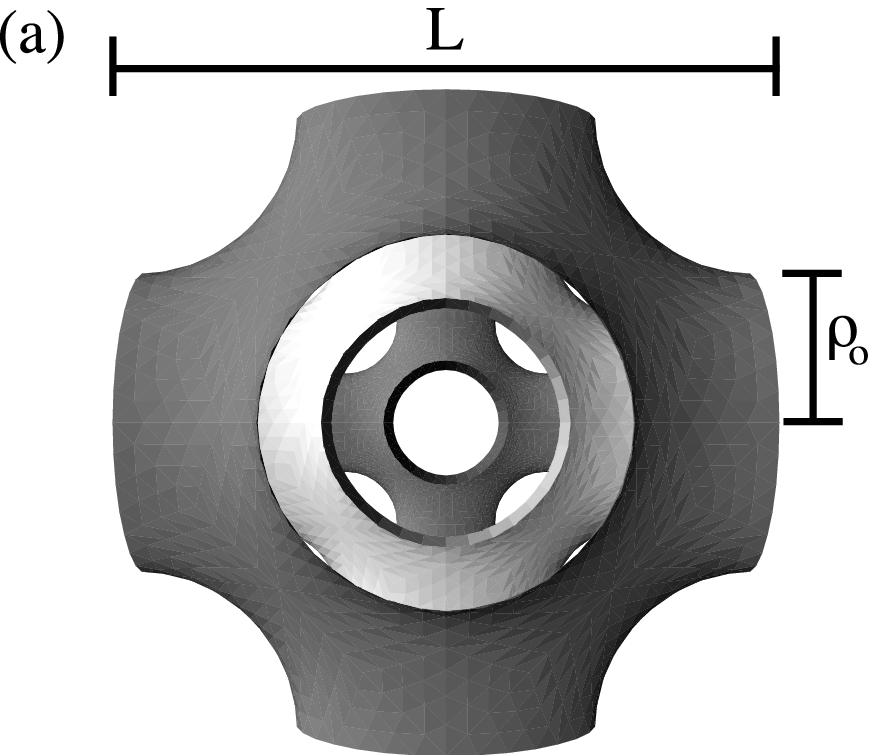, width=2.0in} \epsfig{file=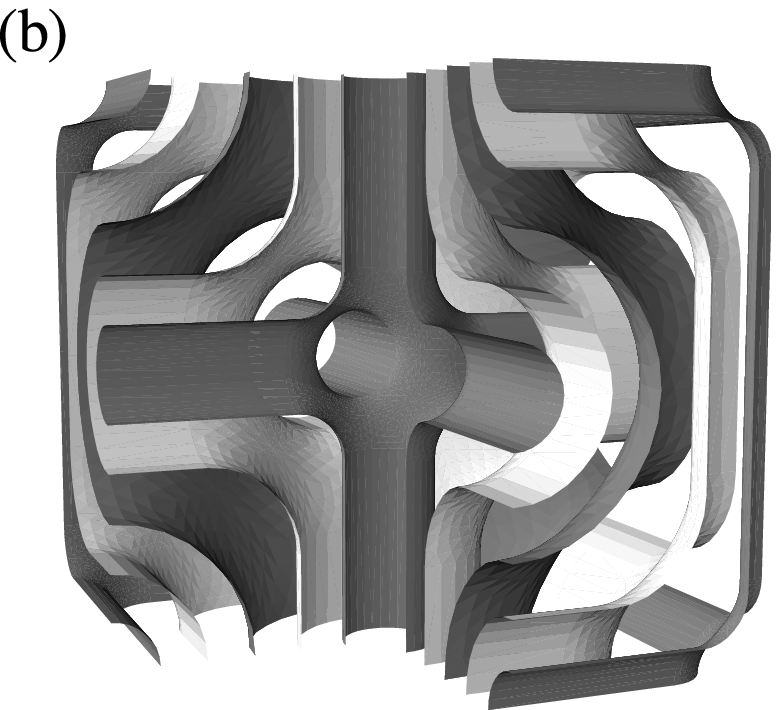, width=2.0in}

\caption{Proposed $P$ surface smectic. In (a) we show the $P$-cell and the interior structure composed of concentric $P$-cells and cylinders.  In (b) we fill in the entire (cubic) unit cell
with eight eighths of the central structure.  We have cut-away one quarter of the volume to show interior smectic layers. }
\label{fig:phasecutout}
\end{figure}

Though we may choose any number of initial space-filling surfaces we illustrate our
approach with a cubic structure.  The Schwartz $P$ surface, or plumber's nightmare, is a triply-periodic minimal surface which has the topology we seek.    Because it is minimal, $H=0$ and
$K\le 0$ everywhere on this surface.
Moreover, because of the mathematical interest in
minimal surfaces, the $P$ surface has a simple parametric representations which proves 
useful in computation.  This surface will be the template within which we fill with ``concentric'' surfaces. 
We refer to the section of the $P$-surface in the unit cell as the $P$-cell and fill it in with concentric, rescaled $P$-cells.  The smaller $P$-cells will no longer 
intersect the walls of the unit-cell.  We attach cylinders to the open ends of the $P$-cells to
fill in the gaps.  The advantage of this construction is that we can easily calculate the
compression energy for the cylinders and the bending energy for the $P$-cells.  Because the inside and the outside of the $P$-surface are identical, our construction self-consistently fills
the entire unit cell and we need only focus on the interior of the $P$-cell -- the remainder of 
the unit cell is filled with eight separate octants of the $P$-cell.  Our approach solves the frustration between compression and bending
by building the smectic phase out of pieces that have no curvature (the $P$-cells) and pieces
which can have uniform spacing (the cylinders) as shown in Fig.~\ref{fig:phasecutout}.  

The energetics of this structure has four  components: the bending energy, the core energy, the saddle-splay energy, and the compression energy.  By our construction there is no bending
energy on the $P$-cells since they are minimal. Thus the curvature energy is nonvanishing only on the
cylindrical sections of the unit-cell.   If $\rho_o$ is the radius of the
largest cylinder, then geometry requires $\rho_o=L/4$ where $L$ is the size of the unit cell.  Since, by construction, the length of the cylindrical region of 
radius $r$ is $2\rho_o-2r$ and $H=1/(2r)$, the bending energy is
\begin{eqnarray}
F_B 
&=& 
24\pi \rho_o K_1 \left( \log\left(\frac{\rho_o}{\rho_C}\right) +
\left(\frac{\rho_C}{\rho_o}-1\right)\right),
\label{eq:fb}
\end{eqnarray}
where $\rho_C$ is the short-distance cutoff.   Inside this cutoff region the smectic order vanishes and there is
an energy penalty proportional to the disordered region.  If the
cutoff is comparable to the molecular scale then, in addition, there is a $+1$ nematic
disclination line down the cylindrical core which contributes to the core energy as well \cite{note}.   Including
the energy of the core region at the intersection of these lines, we have :
\begin{equation}
F_{\rm core} = 24(\rho_o-\rho_C)\varepsilon + \frac{64}{\pi}\rho_C\varepsilon
\end{equation}
where $\varepsilon$ is a line tension.
The saddle-splay energy offsets these positive energy contributions.  As in the traditional chiral
blue phases, the saddle-splay can easily be calculated along the defect lines.  Converting the volume integral (\ref{eq:ssd}) into a surface integral we see that the cylindrical cores of the $P$-cells contribute $-2\pi$ per unit length, and each unit cell has a length $24(\rho_o-\rho_C)$ of such tubes (including the contribution from the other $P$-cell in the unit cell).  Because the integral
over the $P$-cell in the center vanishes, $F_{SS} =  -48\pi a \vert K_{24}\vert N$, where
$N=(\rho_o-\rho_C)/a$ is the number of layers (recall that $K_{24}$ must be negative).  Like the
double-twist tubes in the chiral blue phase, the $P$-cells themselves do not contribute to the saddle-splay though it is their topology that forces line-like topological defects. 
 
The final energetic contribution to the smectic free energy arises from the
nonuniform spacing of the $P$-cells.  By our construction, the part of the smectic which consists of concentric pieces of $P$ surface cannot have uniform spacing across the layers. However, there is an energetically preferred difference in average radius between consecutive layers.  We
can vary the layer spacing to minimize the compression energy. 
To perform the spatial integral we divide our proposed structure into those parts which have cylindrical symmetry and those which are pieces of $P$ surface.
In each region we define our coordinates such that surfaces of constant phase are constants of one variable. In the cylindrical region we use polar coordinates.  Since each $P$-cell is
a minimal surface, it remains minimal when we uniformly rescale it.  Thus for each concentric
$P$-cell, we may define a new variable $\zeta$ through
\begin{equation}
r = \zeta r_\circ \left(\theta, \phi \right),
\label{eq:zetadefine}
\end{equation}
where $r$, $\theta$, and $\phi$ are the usual spherical coordinates.  In the central region
$\zeta$ runs from $0$ to $1$ and so $r=r_\circ(\theta,\phi)$ is the
equation of the $P$-cell. 

If the phase field $\Phi$ describes concentric layers, then $\Phi[\zeta r_\circ(\theta,\phi),\theta,\phi]$ 
must be a constant (independent of $\theta$ and $\phi$) for fixed $\zeta$: therefore
$\Phi[\zeta r_\circ(\theta,\phi),\theta,\phi]$ can only depend on $\zeta$.  However, since the
$(\zeta,\theta,\phi)$ coordinate system is not orthogonal, $\vert\nabb\Phi\vert$ can depend on
all three coordinates.  To calculate $\vert\nabb\Phi\vert$ we note that the layer
normal is parallel to $\nabb\Phi$ and so
$r \partial_r\Phi = {\bf r}\cdot\nabb\Phi = \left({\bf r}\cdot{\bf N}\right)\vert\nabb\Phi\vert$.
But since $\Phi = \Phi(\zeta) = \Phi ( r / r_\circ )$, we also have
$r \partial_r \Phi = (r/r_\circ) \partial_\zeta \Phi = 
\zeta \partial_\zeta \Phi (\zeta)$
and so $({\bf r}\cdot{\bf N})\vert\nabb\Phi\vert = \zeta\partial_\zeta\Phi(\zeta)$ and is thus
constant on any surface of fixed $\zeta$.   Choosing an arbitrary reference direction $\left( \theta_\circ, \phi_\circ \right)$ we have
$\left(\nabb \Phi\right)^2 = p( \theta, \phi ) \Delta ( \zeta )$,
with
\begin{eqnarray}
p( \theta, \phi ) &=&\left[ \frac{{\bf r} (\theta_\circ, \phi_\circ) \cdot {\bf N} (\theta_\circ, \phi_\circ)}
{{\bf r} (\theta, \phi) \cdot {\bf N} (\theta, \phi)}\right]^2\\
\Delta ( \zeta ) &=& \left[\nabb\Phi\right]^2{\Big\vert}_{\zeta,\theta_\circ, \phi_\circ}.
\label{eq:delzetadef}
\end{eqnarray}
Where $p( \theta, \phi )$ is evaluated on a shell of constant $\Phi$.  Note that
$p(\theta,\phi)$ is scale independent and may be evaluated on any of the concentric $P$-cells. 
By choosing the reference direction to be a point on the interface between the $P$-cell region
and the cylindrical region we can define $\zeta  = \rho/\rho_o$ inside each of the six cylindrical regions, where $\rho$ is the radius in cylindrical coordinates. 
This choice allows us to naturally continue $\Delta(\zeta)$ into the
cylindrical regions. The angular and $\zeta$ integrations 
decouple in these coordinates:
\begin{eqnarray}
F_C &=& \frac{B}{2} \rho_o^3 
\int d \zeta  \bigg[ \zeta^2\left( I_{P1} - I_{P2} \Delta (\zeta)  +  I_{P3} \Delta^2 (\zeta)\right) \nonumber\\
&&\qquad+  24 \pi \zeta \left( 1 - \zeta \right) \left( 1 - \Delta (\zeta) \right)^2\bigg] ,
\label{eq:fczeta}
\end{eqnarray}
where the three numerical constants $I_{P1}\approx 72.42$, $I_{P2}\approx 131.48$, and $I_{P3}\approx 68.38$ are moments of $p(\theta,\phi)$ which capture the geometry of the $P$-cell and were calculated with the aid of the the Surface Evolver software package~\cite{Brakke92}.
Since the other contributions to the total free energy do not depend on $\Delta(\zeta)$, we 
minimize $F_C$ by varying the integrand in (\ref{eq:fczeta}) with respect to $\Delta(\zeta)$
and find that the resulting total compression energy in the layered $P$ surface structure is 
$F_C \approx 1.54 B \rho_o^{3} $, with a very weak dependence on the core size $\rho_C$ for
$\rho_C\ll\rho_o$.   

For our proposed smectic structure to be stable against the uniform flat phase, the positive energy contributions from $F_C$, $F_B$, and $F_{\rm core}$ must be compensated by a large negative saddle-splay energy.    Choosing the cutoff $\rho_C$ to be the molecular scale $a$, we minimize the free energy density $\mathcal{F} = (4\rho_o)^{-3}\left[F_C + F_B + F_{\rm core} + F_{SS}\right]$ with respect to $\rho_o$ and find a stable minimum when
\begin{eqnarray}
\nonumber \vert K_{24}\vert&\gtapprox&\varepsilon/\pi, \\
\vert K_{24}\vert &\gg& K_1 N\ln N\nonumber\\
\vert K_{24}\vert&\gg& B\rho_o^2 N
\label{eq:limit}
\end{eqnarray}
The preferred cell size is on the order of:
\begin{equation}
N=\frac{ \rho_o}{a} \approx \frac{3\pi \vert K_{24}\vert + \left[4/\pi - 3/2\right]\varepsilon}{2\pi \vert K_{24}\vert -   \varepsilon}.
\end{equation}
It is worth commenting on the limits in (\ref{eq:limit}).  At the smectic to nematic transition, neither
$K_1$ or $K_{24}$ suffer from anomalous divergences.  Thus this phase is stable only
when $\vert K_{24}\vert$ is exceptionally large.  Numerically minimizing to give a concrete example, we find that when $K_{24}=-500K_1$ and $\varepsilon = 0.99 (2\pi \vert K_{24}\vert)$ that $N\approx 200$.  For this structure to be stable against the flat phase, $B\rho_o^2$ must be smaller than $\sim 80K_1$.  
Since $B= K_1/\lambda^2$ where $\lambda$ is the penetration length,  this requires
that $\lambda > 24a$, {\sl i.e.} the system must be an extreme type II smectic.  Were we to relax the constraint that $H=0$ in the central region, the compression
and curvature could compete and we would expect that $F_C \sim \sqrt{BK_1}\rho_o^2 \sim
B\lambda\rho_o^2$. In this case $\lambda$ could be significantly smaller, though it 
would still scale as $N$.

We note that if this smectic phase is stable then the equivalent nematic phase with $B=0$ would also be stable.  This nematic texture is the splay-analog of the traditional chiral blue phase.  
This splay version of the blue phase
arises through a completely different mechanism than the
traditional cholesteric blue phase.  In the traditional blue phase the defects were kept apart by the
pitch of the cholesteric; here the defects are kept apart by the large energy cost of the line
defects which connect them.  Moreover, the traditional blue phase is stabilized by a large
positive value of $K_{24}$ as opposed to the large negative value required here \cite{sethna}.
Though it is likely that 
the core energy density $\varepsilon$ should be greatly reduced in the case of the nematic since
the smectic condensation energy no longer contributes to the energy, it is possible to have a nematic phase with a periodic lattice of defects.    However, as $\varepsilon$ shrinks, the number of layers approaches a limiting value of $1.5$ and the lattice constant
is comparable to the core size.  This is a pathological limit and, if we were to include
fluctuations, it realizes
the melting of the nematic phase through topological defects.  
The remarkable aspect of both the nematic and the smectic phases presented here is that they exist without any chirality of the molecules. 
Since the stability of this phase requires a careful tuning of the core energy density $\varepsilon$ and $K_{24}$, it can only occur over a small range of temperature.

\begin{figure}

\center

\epsfig{file=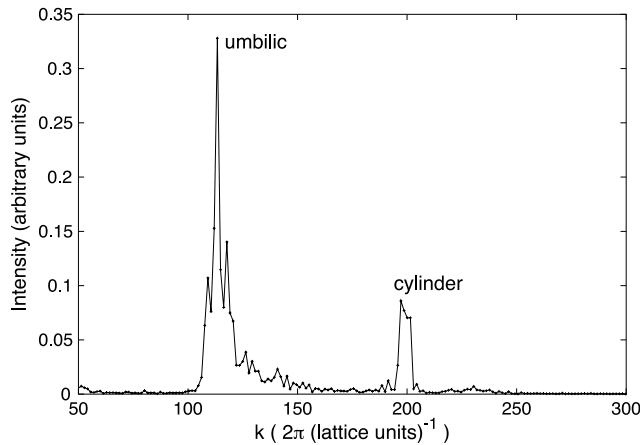, width=3.3in}

\caption{Rotationally averaged form factor for the unit cell with $N=50$.  The umbilic peak corresponds
to the layering along the diagonal of the unit cube.  The cylinder peak arises from the
concentric cylinder regions.}
\label{fig:xray}
\end{figure}

\begin{figure}

\center
\epsfig{file=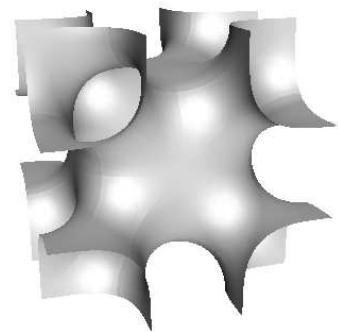, width=2.5in}
\caption{Repeat unit of Schoen's {\sl I-WP} surface. This surface has genus $g=7$ so
$\int K dS = -24\pi$.}
\label{fig:iwp}
\end{figure}

To determine the degree of ordering in our structure, we calculated the powder-averaged
form factor of the unit cell for $N=50$.  Since every line passing through the center of the unit cell intersects
precisely 200 layers, we can infer the degree of ordering along any direction simply from the location of the peak.  Our results are shown in Fig.~\ref{fig:xray}. Not surprisingly, we found a large peak in the directions normal to the eight points at which the $P$ surface is perfectly flat (the umbilics of the $P$ surface). The corresponding direction relative to the unit cell are $(\pm 1, \pm 1, \pm 1)$. We also found weaker peaks along the $(\pm 1,0,0)$, $(0,\pm 1,0)$ and $(0,0,\pm 1)$ 
directions.  These arise from the concentric cylinder regions.  
We note that the eight peaks along
the cell diagonal do not correspond to the observed symmetry of the smectic blue phases, though
this can be changed with a different choice for our unit cell.  However, we have demonstrated
a ``proof-of-principle'' that the construction of splay blue phases is possible.
A better candidate for $BP_{sm}1$ is Schoen's {\sl I-WP} surface, pictured in Figure~\ref{fig:iwp}. This surface is also triply-periodic, though it is flat along the $(\pm 1,0,0)$, $(0,\pm 1,0)$ and $(0,0,\pm 1)$ directions and so in X-ray we would expect the primary peaks along these directions.   Since the {\sl I-WP} surface has genus $g=7$, the integrated surface Gaussian curvature is $-24\pi$ and we expect that the saddle-splay energy would be even more negative than for the $P$-cell.   Calculation of the total energy of an {\sl I-WP} smectic is more difficult than for the $P$ smectic because the region of the cubic cell outside the {\sl I-WP} surface is not identical to the region inside.  Based on the
results here we expect that the compression energy will be relatively unimportant for sufficiently large $\lambda/a$.  

We have outlined a model for smectic blue phases and the splay-analog of the traditional blue phase.  The favorable saddle-splay energy of our structure is sufficient to stabilize it against the flat smectic phase for a realistic range of material parameters which may be encountered near the isotropic transition. The model includes the flexibility to fit previous data on physical smectic blue phases. More importantly, our model presents an entirely new and promising organizational principle for smectic systems.

The authors thank P.A. Heiney, T. C. Lubensky, B. Pansu and T.R. Powers for insightful criticism and A. Finnefrock for useful assistance.  This
work was supported by NSF Grants DMR01-02459 and INT99-10017, and by
a gift from L.J. Bernstein.

\end{document}